\documentclass[letter]{aa} % for the letters 
\usepackage{graphicx,natbib}
\usepackage{txfonts}
\begin{document}
   \title{Multifrequency monitoring of the blazar \object{0716+714} during the 
   GASP-WEBT-AGILE campaign of 2007\thanks{The radio-to-optical data 
   presented in this paper are stored in the WEBT archive; for questions regarding their availability,
   please contact the WEBT President Massimo Villata ({\tt villata@oato.inaf.it}).}}

%   \subtitle{}

   \author{M.~Villata                 \inst{ 1}
   \and   C.~M.~Raiteri               \inst{ 1}
   \and   V.~M.~Larionov              \inst{ 2,3}
   \and   O.~M.~Kurtanidze            \inst{ 4,5,6}
   \and   K.~Nilsson                  \inst{ 7}
   \and   M.~F.~Aller                 \inst{ 8}
   \and   M.~Tornikoski               \inst{ 9}
   \and   A.~Volvach                  \inst{10}
   \and   H.~D.~Aller                 \inst{ 8}
   \and   A.~A.~Arkharov              \inst{ 3}
   \and   U.~Bach                     \inst{11}
   \and   P.~Beltrame                 \inst{12}
   \and   G.~Bhatta                   \inst{13}
   \and   C.~S.~Buemi                 \inst{14}
   \and   M.~B\"ottcher               \inst{15}
   \and   P.~Calcidese                \inst{16}
   \and   D.~Carosati                 \inst{17}
   \and   A.~J.~Castro-Tirado         \inst{18}
   \and   D.~Da~Rio                   \inst{12}
   \and   A.~Di~Paola                 \inst{19}
   \and   M.~Dolci                    \inst{20}
   \and   E.~Forn\'e                  \inst{21}
   \and   A.~Frasca                   \inst{14}
   \and   V.~A.~Hagen-Thorn           \inst{ 2}
   \and   J.~Heidt                    \inst{ 6}
   \and   D.~Hiriart                  \inst{22}
   \and   M.~Jel\'{\i}nek             \inst{18}
   \and   G.~N.~Kimeridze             \inst{ 4}
   \and   T.~S.~Konstantinova         \inst{ 2}
   \and   E.~N.~Kopatskaya            \inst{ 2}
   \and   L.~Lanteri                  \inst{ 1}
   \and   P.~Leto                     \inst{23}
   \and   R.~Ligustri                 \inst{12}
   \and   E.~Lindfors                 \inst{ 7}
   \and   A.~L\"ahteenm\"aki          \inst{ 9}
   \and   E.~Marilli                  \inst{14}
   \and   E.~Nieppola                 \inst{ 9}
   \and   M.~G.~Nikolashvili          \inst{ 4}
   \and   M.~Pasanen                  \inst{ 7}
   \and   B.~Ragozzine                \inst{15}
   \and   J.~A.~Ros                   \inst{21}
   \and   L.~A.~Sigua                 \inst{ 4}
   \and   R.~L.~Smart                 \inst{ 1}
   \and   M.~Sorcia                   \inst{22}
   \and   L.~O.~Takalo                \inst{ 7}
   \and   M.~Tavani                   \inst{24}
   \and   C.~Trigilio                 \inst{14}
   \and   R.~Turchetti                \inst{12}
   \and   K.~Uckert                   \inst{15}
   \and   G.~Umana                    \inst{14}
   \and   S.~Vercellone               \inst{25}
   \and   J.~R.~Webb                  \inst{13}
 }

   \offprints{M.\ Villata}

   \institute{
 % 1
          INAF, Osservatorio Astronomico di Torino, Italy                                                     
 %           2
   \and   Astronomical Institute, St.-Petersburg State University, Russia                                     
 %           3
   \and   Pulkovo Observatory, Russia                                                                         
 %           4
   \and   Abastumani Astrophysical Observatory, Georgia                                                       
 %           5
   \and   Astrophysikalisches Institut Potsdam, Germany                                                       
 %           6
   \and   Landessternwarte Heidelberg-K\"onigstuhl, Germany                                                   
 %           7
   \and   Tuorla Observatory, University of Turku, Finland                                                    
 %           8
   \and   Department of Astronomy, University of Michigan, MI, USA                                            
 %           9
   \and   Mets\"ahovi Radio Obs., Helsinki Univ.\ of Tech.\ TKK, Finland                       
 %          10
   \and   Radio Astronomy Lab.\ of Crimean Astrophysical Obs., Ukraine                            
 %          11
   \and   Max-Planck-Institut f\"ur Radioastronomie, Germany                                                  
 %          12
   \and   Circolo Astrofili Talmassons, Italy                                                                 
 %          13
   \and   SARA Observatory, Florida International University, FL, USA                                         
 %          14
   \and   INAF, Osservatorio Astrofisico di Catania, Italy                                                    
 %          15
   \and   Astrophys.\ Inst., Dept.\ of Phys.\ \& Astron., Ohio Univ., OH, USA              
 %          16
   \and   Oss.\ Astronomico della Regione Autonoma Valle d'Aosta, Italy                                
 %          17
   \and   Armenzano Astronomical Observatory, Italy                                                           
 %          18
   \and   Instituto de Astrof\'{\i}sica de Andaluc\'{\i}a, CSIC, Spain                                        
 %          19
   \and   INAF, Osservatorio Astronomico di Roma, Italy                                                       
 %          20
   \and   INAF, Osservatorio Astronomico di Teramo, Italy                                                     
 %          21
   \and   Agrupaci\'o Astron\`omica de Sabadell, Spain                                                        
 %          22
   \and   Instituto de Astronom\'{\i}a, UNAM, Mexico                                                          
 %          23
   \and   INAF, Istituto di Radioastronomia Sezione di Noto, Italy                                            
 %          24
   \and   INAF, IASF-Roma, Italy                                                                              
 %          25
   \and   INAF, IASF-Milano, Italy                                                                            
 }

   \date{}

  \abstract
  % context heading (optional)
  % {} leave it empty if necessary  
   {} 
  % aims heading (mandatory)
   {Since the CGRO operation in 1991--2000, one of the primary unresolved questions about the blazar $\gamma$-ray emission
has been its possible correlation with the low-energy (in particular optical) emission. To help answer this problem, the 
Whole Earth Blazar Telescope (WEBT) consortium has organized the GLAST-AGILE Support Program (GASP) to provide 
the optical-to-radio monitoring data to be compared with the $\gamma$-ray detections by the AGILE and GLAST satellites.
This new WEBT project started in early September 2007, just before a strong $\gamma$-ray detection of \object{0716+714}
by AGILE.}
  % methods heading (mandatory)
   {We present the GASP-WEBT optical and radio light curves of this blazar obtained in July--November 2007, 
about various AGILE pointings at the source. We construct NIR-to-UV spectral energy distributions (SEDs), 
by assembling GASP-WEBT data together with UV data from the Swift ToO observations of late October.}
  % results heading (mandatory)
   {We observe a contemporaneous optical-radio outburst, which is a rare and interesting
phenomenon in blazars. The shape of the SEDs during the outburst appears peculiarly wavy because of an optical excess and 
a UV drop-and-rise. 
The optical light curve is well sampled during the AGILE pointings, showing prominent and sharp flares.
A future cross-correlation analysis of the optical and AGILE data will shed light on the expected relationship between 
these flares and the $\gamma$-ray events.}
  % conclusions heading (optional), leave it empty if necessary 
  {}

  \keywords{galaxies: active -- galaxies: BL Lacertae objects:
    general -- galaxies: BL Lacertae objects: individual:
    \object{S5 0716+71} -- galaxies: jets}
%   \titlerunning{}

   \maketitle
%
%________________________________________________________________

\section{Introduction}

   \begin{figure*}
   \sidecaption
   \includegraphics[width=13cm]{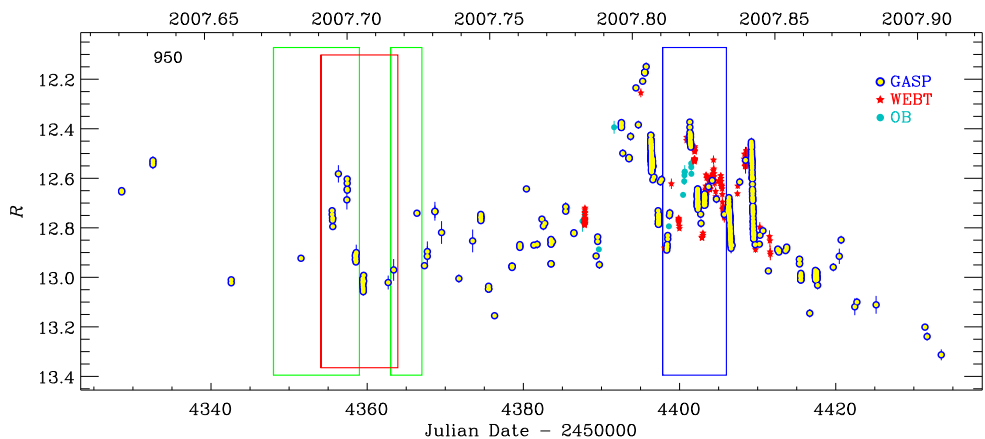}
%% PLEASE DON'T CHANGE THE FIGURE SIZE [width=13cm] AND ITS LOCATION IN THE TEXT, IF POSSIBLE
      \caption{$R$-band light curve of \object{0716+714} in August--November 2007, composed with data from the 
GASP (blue, yellow-filled circles) and from other WEBT telescopes (red stars); the cyan filled circles 
labelled ``OB" (i.e.\ ``other bands") represent GASP $J$-band data converted to $R$ band. 
The boxes indicate various periods of AGILE observations, as explained in the text.}
      \label{fig1}
   \end{figure*}

The Compton Gamma Ray Observatory (CGRO, 1991--2000) discovered that blazars can be strong $\gamma$-ray emitters, 
and the third EGRET catalog \citep{har99} contains 66 high-confidence and 27 low-confidence identifications of  
blazars at energies greater than 100 MeV \citep[see also][]{mat01}.
The $\gamma$-ray emission from blazars is usually ascribed to inverse-Compton scattering of soft photons by 
synchrotron-emitting relativistic electrons in the plasma jet. 
According to synchrotron-self-Compton (SSC) models, the source of soft photons is the synchrotron process itself, 
which produces the radio-to-optical 
(radio-to-X-rays for the high-energy-peaked BL Lac objects) non-thermal emission. 
In contrast, in external-Compton (EC) models,
seed photons come from outside the jet, in particular from the accretion disc or the broad line region. 
Finally, $\gamma$-ray emission might also originate from particle cascades.
Different models make different predictions on the multifrequency behaviour of the source, 
which must be tested on observational grounds. 
A few multiwavelength observations were organized during the CGRO operation,
whose analysis provided some indication of optical-$\gamma$ correlation \citep[e.g.][]{blo97,har01}.

A deeper insight into the high-energy emission production mechanism can be attained only by 
organizing ground-based multiwavelength campaigns coordinated with the satellite observations, in which
a large number of telescopes at different longitudes is involved, to obtain continuous monitoring.
Indeed, this was one of the main motivations that led to the birth of the Whole Earth Blazar Telescope 
(WEBT)\footnote{{\tt http://www.oato.inaf.it/blazars/webt/}\\ see e.g.\ \citet{vil04a,vil06,rai06b,rai07b}.} 
in 1997, during the operation of CGRO. 
In ten years of activity, the WEBT consortium has proved its capability to obtain high-precision 
and well-sampled light curves in the optical, near-IR, and radio bands, during more than twenty monitoring campaigns.

More recently, following the launch of the $\gamma$-ray satellites 
Astro-rivelatore Gamma a Immagini LEggero (AGILE) and Gamma-ray Large Area Space Telescope (GLAST),
the WEBT has started a new project: the GLAST-AGILE Support Program (GASP). Its primary aim is to provide 
long-term continuous monitoring of a list of selected $\gamma$-loud blazars, during the operation of these two satellites, 
by means of some fifteen WEBT telescopes.
AGILE was launched on April 23, 2007, and one of its first detections of extragalactic objects was \object{S5 0716+71}, 
in mid September \citep{giu07}.

The blazar S5 0716+71 is a BL Lac object of still unknown redshift. 
It is famous for its strong variability, on both long and short (intraday) timescales \citep[e.g.][]{wag96,ghi97,rai03}, and it
was the target of two WEBT campaigns in 1999 and 2003--2004 \citep{vil00,ost06}.

The detection by AGILE occurred at the beginning of the optical observing season, as well as at the start 
of the GASP activity.
The data collected by the GASP showed that an optical flare had occurred in the same period \citep{car07}.
About one month later, the GASP observed a new very bright phase, triggering new AGILE as well as Swift observations. 
The alert was also spread among the other WEBT observers to improve the sampling.

In this paper, we report on the GASP-WEBT multifrequency observations and UV Swift data in July--November 2007, 
around the AGILE pointings at the source.

\section{Optical-to-radio observations by the GASP and WEBT}

   \begin{figure}
   \resizebox{\hsize}{!}{\includegraphics{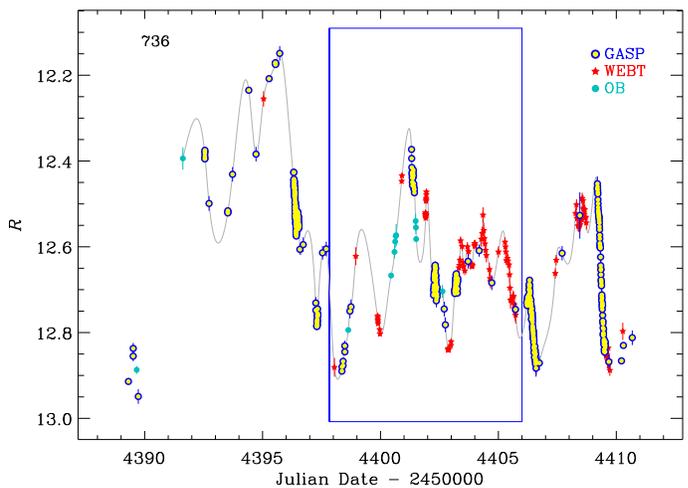}}
%% PLEASE DON'T CHANGE THE FIGURE SIZE (\resizebox{\hsize}{!}) AND ITS LOCATION IN THE TEXT, IF POSSIBLE
      \caption{Enlargement of the best-sampled part of the $R$-band light curve shown in Fig.\ \ref{fig1}, 
during the bright and active phase of October--November 2007. Symbols and box as in Fig.\ \ref{fig1}. 
The grey curve represents a cubic spline interpolation through the 30-min binned data.}
         \label{fig2}
   \end{figure}

We calibrated the optical $R$-band magnitudes with respect to Stars 2, 3, 5, and 6 from the photometric sequence by \citet{vil98b}.
Figure \ref{fig1} shows the $R$-band data collected by the GASP and WEBT in August--November 2007.
We added some $J$-band data converted to the $R$-band (adopting a mean colour index $R-J=1.46$), to improve the sampling in 
the critical periods.
In total, we assembled 950 data points, acquired mainly from the GASP observatories: Abastumani, Armenzano, Crimean, L'Ampolla, 
Roque de los Muchachos (KVA), Sabadell, St.\ Petersburg, Talmassons, Torino, Valle d'Aosta, and Campo Imperatore, this last 
providing $JHK$ data. 
Other participating WEBT observatories were BOOTES-2\footnote{30 cm telescope, see \citet{cas04}.}, Kitt Peak (MDM and SARA), 
and San Pedro Martir, supplying data about the October--November outburst.

In the figure, the green boxes indicate the periods of the first AGILE pointings, from September 
4 to 23\footnote{The AGILE pointing periods shown in Fig.\ \ref{fig1} represent the dates when \object{0716+714} was 
in the AGILE field of view at less than 40\degr\ from the axis.
During other pointings the source was more off-axis, where the data analysis is currently more uncertain.}, 
including the strong ($\sim 8 \sigma$) $\gamma$-ray detection of September 10--20 reported by \citet{giu07}, 
which is represented by the red box. In the same period, we observed a noticeable optical flare \citep[see also][]{car07}.
Then, after a rather variable phase, in mid October the optical flux started to rise, reaching a peak 
of $R=12.15$\footnote{This high brightness level is comparable to those reached in the recurrent outbursts of 2004--2007, 
which represent the brightest states ever observed.} on October 22.2\footnote{On the same date the source was seen bright 
in the AGILE field of view, even if it was rather off-axis.}. 
This triggered intensive observations at all wavelengths and in particular AGILE and Swift ToO pointings.
Because of a sudden drop of 0.73 mag in 2.3 days, the optical level was relatively low at the start of the AGILE observations 
on October 24.3. 
AGILE remained on the source until November 1.5, and during this period (blue box) \object{0716+714} showed again strong variability and 
bright phases. 
The analysis of the $\gamma$-ray data taken by AGILE is still underway and will be presented in a forthcoming paper 
(Chen et al., in preparation).

   \begin{figure}
   \resizebox{\hsize}{!}{\includegraphics{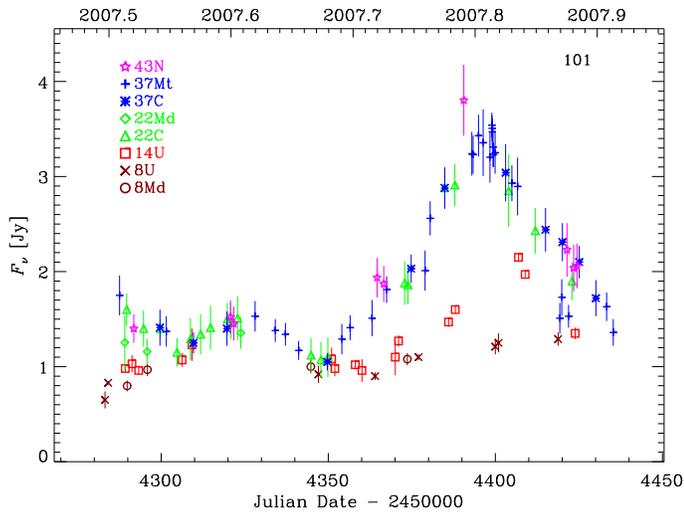}}
%% PLEASE DON'T CHANGE THE FIGURE SIZE (\resizebox{\hsize}{!}) AND ITS LOCATION IN THE TEXT, IF POSSIBLE
      \caption{Radio data from 8 to 43 GHz of 0716+714 in July--November 2007. Different symbols refer to different 
frequencies and observatories, as explained in the text.}
         \label{fig3}
   \end{figure}

   \begin{figure}
   \resizebox{\hsize}{!}{\includegraphics{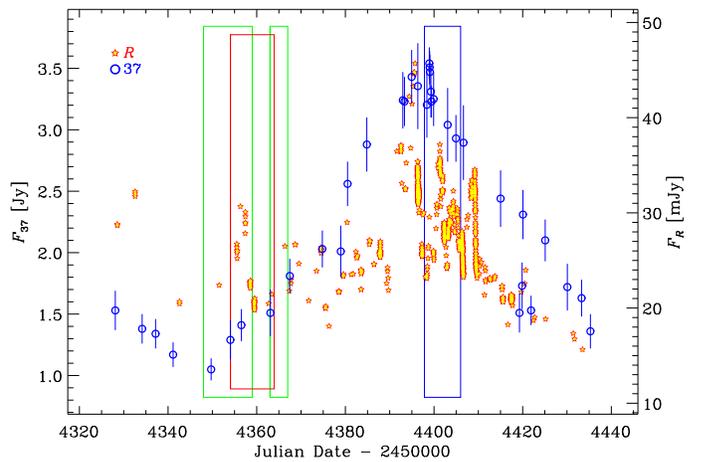}}
%% PLEASE DON'T CHANGE THE FIGURE SIZE (\resizebox{\hsize}{!}) AND ITS LOCATION IN THE TEXT, IF POSSIBLE
      \caption{37 GHz light curve of 0716+714 in August--November 2007 (blue circles) compared to the $R$-band flux densities in 
the same period (red, yellow-filled stars), rescaled to have the optical and radio maxima at the same level.}
         \label{fig4}
   \end{figure}

The October--November optical outburst is better displayed in Fig.\ \ref{fig2}, where we traced a cubic spline interpolation through 
the 30-min binned light curve, to guide the eye along the fast and complex variations. 
In particular, we notice the fall of 0.39 mag in 7.1 hours on $\rm JD = 2454409.20$--9.49 (November 4), during which 
the steepest slopes match the typical values, $\sim 0.002$ mag per minute, estimated by \citet[see also
\citealt{ost06} and references therein]{vil00}.

The subsequent period (see Fig.\ \ref{fig1}) is characterized by an apparently smooth brightness decrease\footnote{The fall continued 
also beyond the period shown in the figure, and led the source to reach $R \sim 15$ in mid December, 
comparable to the historical minima of 1996 and 2000 \citep{rai03}.}.

In Fig.\ \ref{fig3}, we show the radio data taken by the GASP and WEBT in July--November 2007 in the 8--43 GHz frequency range. 
The 8 GHz data come from Medicina and UMRAO (8Md and 8U in the figure legend, respectively); 
14.5 GHz data are from UMRAO (14U); 22 GHz observations were carried out at the Crimean (RT-22) and Medicina observatories (22C and 22Md); 
37 GHz ones at the Crimean (RT-22) and Mets\"ahovi telescopes (37C and 37Mt), while 43 GHz data are from Noto (43N). 
The light curves show a relatively quiet initial phase with a slightly inverted spectrum. 
Around $\rm JD=2454350$ the higher-frequency (22--43 GHz) fluxes started to increase, 
reaching a maximum level of about 3.5 Jy at 37 GHz\footnote{This radio peak represents an unusually bright level, though it 
appears modest when compared with the 2003 radio outburst ($\sim 6$ Jy at 37 GHz, \citealp[see e.g.][]{ter05,nie07}).} more or less 
at the same time as the optical peak.
At 14.5 GHz, the increase appears less pronounced, and delayed by about 20 days.
The 8 GHz data show only a marginal flux enhancement, and no clear signature of the outburst 
is seen in the rather sparse 5 GHz data, not shown in the figure.

Figure \ref{fig4} displays the 37 GHz and optical $R$-band flux densities in August--November 2007 together.
The optical and radio outbursts appear roughly contemporaneous.
However, when seen in detail, the two events show a different behaviour.
In general, the optical flux presents stronger and faster variations, whereas the radio flux rises and falls
in a much smoother way.
Optical and radio fluxes appear to peak almost at the same time, but while the optical flux drops in a few hours, 
the high radio state lasts for a few days.
The discrete correlation function (DCF) analysis, applied to the optical and radio light curves, 
yields a radio delay of about 4 days. 
Other main differences in flux behaviour are the absence of radio counterparts to the August and September optical flares, 
and the unusual fact that the radio flux achieves a very bright state well before the optical flux.
Thus, due to these differences, the 4-day radio delay, though statistically significant ($\rm DCF > 1$), must
be taken with caution, since it mainly reflects the time separation of the detected optical and radio maxima, 
with their different sampling.

A contemporaneous optical-radio event such as that detected in October--November has sometimes been observed in blazars
\citep[see e.g.][]{tor94a,tor94b}. However, most studies on radio-optical correlation have shown that the radio 
events lag behind the optical ones by several weeks or months \citep[see e.g.][]{tor94b,cle95,vil04b,vil07,rai08}.
In the case of 0716+714, \citet{rai03} noticed that in the 1994--2001 observing period, major optical outbursts may have 
modest radio counterparts. 
Thus, the optically-emitting jet region is sometimes not completely opaque to the high radio frequencies.
In particular, it is quite transparent to them in the case of the 2007 outburst, while lower frequencies 
are at least partially absorbed, as the 14.5 GHz delay and the strongly-inverted spectrum indicate.

\section{Swift-UVOT observations and NIR-to-UV SEDs}

Besides the AGILE ToO pointing, the optical brightening of the source also triggered observations by the Swift satellite, 
which started on October 23 and lasted until mid November 2007. The UltraViolet and Optical Telescope (UVOT) 
instrument acquired data in all of its optical ($VBU$) and ultraviolet (UV$W1$, UV$M2$, UV$W2$) bands. 
These data were processed with the {\tt uvotmaghist} task of the HEASOFT 6.3 package. 
The source counts were extracted from a circular region of 5 arcsec radius and the background counts 
from a surrounding annulus with 8 and 18 arcsec radii. A 0.1 mag error was assumed, to take into account both systematic and 
statistical errors.

\begin{figure}
   \resizebox{\hsize}{!}{\includegraphics{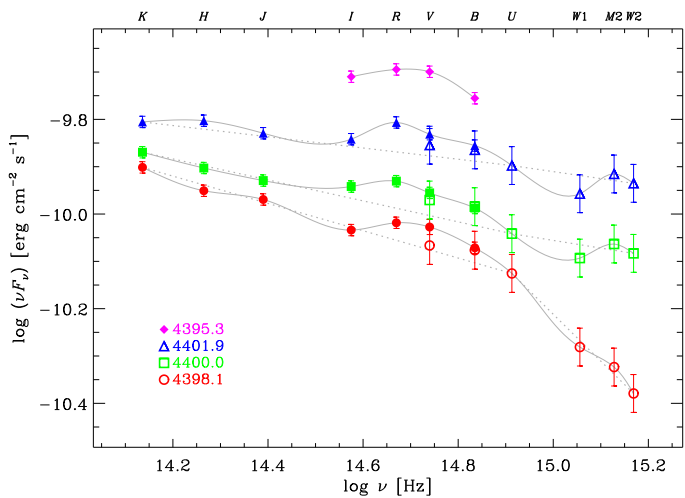}}
%% PLEASE DON'T CHANGE THE FIGURE SIZE (\resizebox{\hsize}{!}) AND ITS LOCATION IN THE TEXT, IF POSSIBLE
      \caption{Spectral energy distributions (SEDs) of 0716+714 in late October 2007, 
constructed with $W2\,M2\,W1\,U\,B\,V$ data from Swift-UVOT (empty symbols) and ground-based GASP-WEBT $BVRIJHK$ data (filled symbols). 
The numbers in the legend refer to the Julian Dates of the observations. 
The dotted lines connect the $K$, $U$, and $W2$ data; together with the cubic splines through the spectra, they
help to better distinguish the main features of the SEDs (see text).}
         \label{fig5}
   \end{figure}

Figure \ref{fig5} shows the source SEDs built with $V$ to UV$W2$ data from UVOT and $BVRIJHK$ data from GASP-WEBT, at three epochs 
characterized by different brightness levels\footnote{Magnitudes were corrected for Galactic extinction adopting $A_B=0.132$ and 
deriving the values in the other bands according to \citet{car89}.}.
In the SEDs at $\rm JD=4400.0$ and $\rm JD=4401.9$, the GASP-WEBT optical data are strictly contemporaneous to the UVOT data, 
while the ground-based optical data in the lowest-state SED at $\rm JD=4398.1$ come from another date when the source was at 
the same optical level.
In addition, the $JHK$ data are not strictly contemporaneous to the UVOT data, but again taken at the same brightness levels.
Finally, the highest-state optical SED (close to the optical peak) is plotted for comparison.

The SEDs show two main features: the bluer-when-brighter behaviour (especially in the UV band), 
which is typical of BL Lac objects, and the wavy shape of the spectra.
In particular, we notice the bumpy aspect of the optical part and the dip in the UV$W1$ band.
A deep investigation/discussion on these SED features is beyond the scope of this letter.
We note that the dip in the UV$W1$ band resembles an analogous feature found in the optical--UV SEDs of AO 0235+16 
by \citet{rai08}, who interpreted it as anomalous absorption, or as a UV excess at higher frequencies. However, in our case
the drop, though systematic, is within the uncertainties, and might be due to calibration problems.

\section{Conclusions}

We have reported on the multifrequency behaviour of the blazar 0716+714 in July--November 2007, about the AGILE pointings at the source. 
Radio-to-optical data were taken by the GASP, supported by other WEBT observatories, in particular during the most critical period of the 
October--November outburst, when ToO observations by AGILE and Swift were triggered as a consequence of the detected optical brightening.

For the first time, a contemporaneous optical-radio outburst was followed in detail, 
showing both the similarities and the differences between the optical and radio behaviours.

The NIR-to-UV SEDs constructed during the most active phase of late October with both ground-based GASP-WEBT and space Swift-UVOT data 
also show unexpected features, such as the optical excess and the UV drop-and-rise.

In particular, we have assembled a detailed optical light curve during the AGILE ToO pointing of October 24 -- November 1, and 
detected a prominent optical flare in correspondence with the strong $\gamma$-ray detection of mid September.
This optical information will be very useful when the $\gamma$-ray data are completely analysed, because it will be possible 
to search for correlations and time delays between optical flares and $\gamma$-ray emission.

\begin{acknowledgements}
We thank A.\ Capetti for useful discussions.
We acknowledge the use of public data from the Swift data archive.
This research has made use of data obtained through the High Energy Astrophysics Science Archive Research Center Online Service, 
provided by the NASA/Goddard Space Flight Center.
AZT-24 observations at Campo Imperatore are made within an agreement between Pulkovo, Rome and Teramo observatories.
This research has made use of data from the University of Michigan Radio Astronomy Observatory,
which is supported by the National Science Foundation and by funds from the University of Michigan.
This work is partly based on observations with the Medicina and Noto radio telescopes operated
by INAF -- Istituto di Radioastronomia.
The Torino team acknowledges financial support by the Italian Space Agency through contract 
ASI-INAF I/088/06/0 for the Study of High-Energy Astrophysics. 
The Mets\"ahovi team acknowledges the support from the Academy of Finland.
JH acknowledges support by the German Science Foundation (DFG) through SFB 439.
MJ would like to thank the Spanish Ministry of Education
and Science for the support via grants AP2003-1407, ESP2002-04124-C03-01, and AYA2004-01515 (+ FEDER funds).
\end{acknowledgements}


\begin{thebibliography}{29}
\expandafter\ifx\csname natexlab\endcsname\relax\def\natexlab#1{#1}\fi

\bibitem[{{Bloom} {et~al.}(1997){Bloom}, {Bertsch}, {Hartman}, {Sreekumar},
  {Thompson}, {Balonek}, {Beckerman}, {Davis}, {Whitman}, {Miller}, {Nair},
  {Roberts}, {Tosti}, {Massaro}, {Nesci}, {Maesano}, {Montagni}, {Jang},
  {Bock}, {Dietrich}, {Herter}, {Otterbein}, {Pfeiffer}, {Seitz}, \&
  {Wagner}}]{blo97}
{Bloom}, S.~D., {Bertsch}, D.~L., {Hartman}, R.~C., {et~al.} 1997, \apjl, 490,
  L145

\bibitem[{{Cardelli} {et~al.}(1989){Cardelli}, {Clayton}, \& {Mathis}}]{car89}
{Cardelli}, J.~A., {Clayton}, G.~C., \& {Mathis}, J.~S. 1989, \apj, 345, 245

\bibitem[{{Carosati} {et~al.}(2007){Carosati}, {Larionov}, {Larionova}, {Ros},
  {Villata}, \& {Raiteri}}]{car07}
{Carosati}, D., {Larionov}, V.~M., {Larionova}, L., {et~al.} 2007, The
  Astronomer's Telegram, 1223, 1

\bibitem[{{Castro-Tirado} {et~al.}(2004){Castro-Tirado}, {Jel{\'{\i}}nek},
  {Mateo Sanguino}, {de Ugarte Postigo}, \& {the BOOTES team}}]{cas04}
{Castro-Tirado}, A.~J., {Jel{\'{\i}}nek}, M., {Mateo Sanguino}, T.~J., {de
  Ugarte Postigo}, A., \& {the BOOTES team}. 2004, Astronomische Nachrichten,
  325, 679

\bibitem[{{Clements} {et~al.}(1995){Clements}, {Smith}, {Aller}, \&
  {Aller}}]{cle95}
{Clements}, S.~D., {Smith}, A.~G., {Aller}, H.~D., \& {Aller}, M.~F. 1995, \aj,
  110, 529

\bibitem[{{Ghisellini} {et~al.}(1997){Ghisellini}, {Villata}, {Raiteri},
  {Bosio}, {de Francesco}, {Latini}, {Maesano}, {Massaro}, {Montagni}, {Nesci},
  {Tosti}, {Fiorucci}, {Pian}, {Maraschi}, {Treves}, {Comastri}, \&
  {Mignoli}}]{ghi97}
{Ghisellini}, G., {Villata}, M., {Raiteri}, C.~M., {et~al.} 1997, \aap, 327, 61

\bibitem[{{Giuliani} {et~al.}(2007){Giuliani}, {Vercellone}, {Chen},
  {Mereghetti}, {Pellizzoni}, {Perotti}, {Fornari}, {Fiorini}, {Caraveo},
  {Zambra}, {Bulgarelli}, {Gianotti}, {Trifoglio}, {Cocco}, {Labanti},
  {Fuschino}, {Marisaldi}, {Galli}, {Tavani}, {Pucella}, {D'Ammando},
  {Vittorini}, {Costa}, {Feroci}, {Donnarumma}, {Pacciani}, {Monte},
  {Lazzarotto}, {Soffitta}, {Evangelista}, {Lapshov}, {Rapisarda}, {Argan},
  {Trois}, {Paris}, {Barbiellini}, {Longo}, {Picozza}, {Morselli}, {Prest},
  {Vallazza}, {Lipari}, {Zanello}, {Mauri}, {Giommi}, {Pittori}, {Antonelli},
  {Gasparrini}, {Cutini}, {Verrecchia}, \& {Salotti}}]{giu07}
{Giuliani}, A., {Vercellone}, S., {Chen}, A., {et~al.} 2007, The Astronomer's
  Telegram, 1221, 1

\bibitem[{{Hartman} {et~al.}(1999){Hartman}, {Bertsch}, {Bloom}, {Chen},
  {Deines-Jones}, {Esposito}, {Fichtel}, {Friedlander}, {Hunter}, {McDonald},
  {Sreekumar}, {Thompson}, {Jones}, {Lin}, {Michelson}, {Nolan}, {Tompkins},
  {Kanbach}, {Mayer-Hasselwander}, {M{\"u}cke}, {Pohl}, {Reimer}, {Kniffen},
  {Schneid}, {von Montigny}, {Mukherjee}, \& {Dingus}}]{har99}
{Hartman}, R.~C., {Bertsch}, D.~L., {Bloom}, S.~D., {et~al.} 1999, \apjs, 123,
  79

\bibitem[{{Hartman} {et~al.}(2001){Hartman}, {Villata}, {Balonek}, {Bertsch},
  {Bock}, {B{\"o}ttcher}, {Carini}, {Collmar}, {De Francesco}, {Ferrara},
  {Heidt}, {Kanbach}, {Katajainen}, {Koskimies}, {Kurtanidze}, {Lanteri},
  {Lawson}, {Lin}, {Marscher}, {McFarland}, {McHardy}, {Miller},
  {Nikolashvili}, {Nilsson}, {Noble}, {Nucciarelli}, {Ostorero}, {Pursimo},
  {Raiteri}, {Rekola}, {Savolainen}, {Sillanp{\"a}{\"a}}, {Smale}, {Sobrito},
  {Takalo}, {Thompson}, {Tosti}, {Wagner}, \& {Wilson}}]{har01}
{Hartman}, R.~C., {Villata}, M., {Balonek}, T.~J., {et~al.} 2001, \apj, 558,
  583

\bibitem[{{Mattox} {et~al.}(2001){Mattox}, {Hartman}, \& {Reimer}}]{mat01}
{Mattox}, J.~R., {Hartman}, R.~C., \& {Reimer}, O. 2001, \apjs, 135, 155

\bibitem[{{Nieppola} {et~al.}(2007){Nieppola}, {Tornikoski},
  {L{\"a}hteenm{\"a}ki}, {Valtaoja}, {Hakala}, {Hovatta}, {Kotiranta},
  {Nummila}, {Ojala}, {Parviainen}, {Ranta}, {Saloranta}, {Torniainen}, \&
  {Tr{\"o}ller}}]{nie07}
{Nieppola}, E., {Tornikoski}, M., {L{\"a}hteenm{\"a}ki}, A., {et~al.} 2007,
  \aj, 133, 1947

\bibitem[{{Ostorero} {et~al.}(2006){Ostorero}, {Wagner}, {Gracia}, {Ferrero},
  {Krichbaum}, {Britzen}, {Witzel}, {Nilsson}, {Villata}, {Bach}, {Barnaby},
  {Bernhart}, {Carini}, {Chen}, {Chen}, {Ciprini}, {Crapanzano}, {Doroshenko},
  {Efimova}, {Emmanoulopoulos}, {Fuhrmann}, {Gabanyi}, {Giltinan},
  {Hagen-Thorn}, {Hauser}, {Heidt}, {Hojaev}, {Hovatta}, {Hroch}, {Ibrahimov},
  {Impellizzeri}, {Ivanidze}, {Kachel}, {Kraus}, {Kurtanidze},
  {L{\"a}hteenm{\"a}ki}, {Lanteri}, {Larionov}, {Lin}, {Lindfors}, {Munz},
  {Nikolashvili}, {Nucciarelli}, {O'Connor}, {Ohlert}, {Pasanen}, {Pullen},
  {Raiteri}, {Rector}, {Robb}, {Sigua}, {Sillanp{\"a}{\"a}}, {Sixtova},
  {Smith}, {Strub}, {Takahashi}, {Takalo}, {Tapken}, {Tartar}, {Tornikoski},
  {Tosti}, {Tr{\"o}ller}, {Walters}, {Wilking}, {Wills}, {Agudo}, {Aller},
  {Aller}, {Angelakis}, {Klare}, {K{\"o}rding}, {Strom}, {Ter{\"a}sranta},
  {Ungerechts}, \& {Vila-Vilar{\'o}}}]{ost06}
{Ostorero}, L., {Wagner}, S.~J., {Gracia}, J., {et~al.} 2006, \aap, 451, 797

\bibitem[{{Raiteri} {et~al.}(2006){Raiteri}, {Villata}, {Kadler}, {Ibrahimov},
  {Kurtanidze}, {Larionov}, {Tornikoski}, {Boltwood}, {Lee}, {Aller}, {Romero},
  {Aller}, {Araudo}, {Arkharov}, {Bach}, {Barnaby}, {Berdyugin}, {Buemi},
  {Carini}, {Carosati}, {Cellone}, {Cool}, {Dolci}, {Efimova}, {Fuhrmann},
  {Hagen-Thorn}, {Holcomb}, {Ilyin}, {Impellizzeri}, {Ivanidze}, {Kapanadze},
  {Kerp}, {Konstantinova}, {Kovalev}, {Kovalev}, {Kraus}, {Krichbaum},
  {L{\"a}hteenm{\"a}ki}, {Lanteri}, {Leto}, {Lindfors}, {Mattox}, {Napoleone},
  {Nikolashvili}, {Nilsson}, {Ohlert}, {Papadakis}, {Pasanen}, {Poteet},
  {Pursimo}, {Ros}, {Sigua}, {Smith}, {Takalo}, {Trigilio}, {Tr{\"o}ller},
  {Umana}, {Ungerechts}, {Walters}, {Witzel}, \& {Xilouris}}]{rai06b}
{Raiteri}, C.~M., {Villata}, M., {Kadler}, M., {et~al.} 2006, \aap, 459, 731

\bibitem[{{Raiteri} {et~al.}(2008){Raiteri}, {Villata}, {Larionov}, {Aller},
  {Bach}, \& {et al.}}]{rai08}
{Raiteri}, C.~M., {Villata}, M., {Larionov}, V.~M., {et~al.} 2008, ArXiv
  e-prints, 8011236

\bibitem[{{Raiteri} {et~al.}(2007){Raiteri}, {Villata}, {Larionov}, {Pursimo},
  {Ibrahimov}, {Nilsson}, {Aller}, {Kurtanidze}, {Foschini}, {Ohlert},
  {Papadakis}, {Sumitomo}, {Volvach}, {Aller}, {Arkharov}, {Bach}, {Berdyugin},
  {B{\"o}ttcher}, {Buemi}, {Calcidese}, {Charlot}, {Delgado S{\'a}nchez}, {di
  Paola}, {Djupvik}, {Dolci}, {Efimova}, {Fan}, {Forn{\'e}}, {Gomez}, {Gupta},
  {Hagen-Thorn}, {Hooks}, {Hovatta}, {Ishii}, {Kamada}, {Konstantinova},
  {Kopatskaya}, {Kovalev}, {Kovalev}, {L{\"a}hteenm{\"a}ki}, {Lanteri}, {Le
  Campion}, {Lee}, {Leto}, {Lin}, {Lindfors}, {Mingaliev}, {Mizoguchi},
  {Nicastro}, {Nikolashvili}, {Nishiyama}, {{\"O}stman}, {Ovcharov},
  {P{\"a}{\"a}kk{\"o}nen}, {Pasanen}, {Pian}, {Rector}, {Ros}, {Sadakane},
  {Selj}, {Semkov}, {Sharapov}, {Somero}, {Stanev}, {Strigachev}, {Takalo},
  {Tanaka}, {Tavani}, {Torniainen}, {Tornikoski}, {Trigilio}, {Umana},
  {Vercellone}, {Valcheva}, {Volvach}, \& {Yamanaka}}]{rai07b}
{Raiteri}, C.~M., {Villata}, M., {Larionov}, V.~M., {et~al.} 2007, \aap, 473,
  819

\bibitem[{{Raiteri} {et~al.}(2003){Raiteri}, {Villata}, {Tosti}, {Nesci},
  {Massaro}, {Aller}, {Aller}, {Ter{\"a}sranta}, {Kurtanidze}, {Nikolashvili},
  {Ibrahimov}, {Papadakis}, {Krichbaum}, {Kraus}, {Witzel}, {Ungerechts},
  {Lisenfeld}, {Bach}, {Cim{\`o}}, {Ciprini}, {Fuhrmann}, {Kimeridze},
  {Lanteri}, {Maesano}, {Montagni}, {Nucciarelli}, \& {Ostorero}}]{rai03}
{Raiteri}, C.~M., {Villata}, M., {Tosti}, G., {et~al.} 2003, \aap, 402, 151

\bibitem[{{Ter{\"a}sranta} {et~al.}(2005){Ter{\"a}sranta}, {Wiren}, {Koivisto},
  {Saarinen}, \& {Hovatta}}]{ter05}
{Ter{\"a}sranta}, H., {Wiren}, S., {Koivisto}, P., {Saarinen}, V., \&
  {Hovatta}, T. 2005, \aap, 440, 409

\bibitem[{{Tornikoski} {et~al.}(1994{\natexlab{a}}){Tornikoski}, {Valtaoja},
  {Ter{\"a}sranta}, \& {Okyudo}}]{tor94a}
{Tornikoski}, M., {Valtaoja}, E., {Ter{\"a}sranta}, H., \& {Okyudo}, M.
  1994{\natexlab{a}}, \aap, 286, 80

\bibitem[{{Tornikoski} {et~al.}(1994{\natexlab{b}}){Tornikoski}, {Valtaoja},
  {Ter{\"a}sranta}, {Smith}, {Nair}, {Clements}, \& {Leacock}}]{tor94b}
{Tornikoski}, M., {Valtaoja}, E., {Ter{\"a}sranta}, H., {et~al.}
  1994{\natexlab{b}}, \aap, 289, 673

\bibitem[{{Villata} {et~al.}(2000){Villata}, {Mattox}, {Massaro}, {Nesci},
  {Catalano}, {Frasca}, {Raiteri}, {Sobrito}, {Tosti}, {Nucciarelli}, {Takalo},
  {Sillanp{\"a}{\"a}}, {Karttunen}, {Maesano}, {Marilli}, {Ostorero},
  {Piironen}, \& {Sclavi}}]{vil00}
{Villata}, M., {Mattox}, J.~R., {Massaro}, E., {et~al.} 2000, \aap, 363, 108

\bibitem[{{Villata} {et~al.}(2004{\natexlab{a}}){Villata}, {Raiteri}, {Aller},
  {Aller}, {Ter{\"a}sranta}, {Koivula}, {Wiren}, {Kurtanidze}, {Nikolashvili},
  {Ibrahimov}, {Papadakis}, {Tosti}, {Hroch}, {Takalo}, {Sillanp{\"a}{\"a}},
  {Hagen-Thorn}, {Larionov}, {Schwartz}, {Basler}, {Brown}, \&
  {Balonek}}]{vil04b}
{Villata}, M., {Raiteri}, C.~M., {Aller}, H.~D., {et~al.} 2004{\natexlab{a}},
  \aap, 424, 497

\bibitem[{{Villata} {et~al.}(2007){Villata}, {Raiteri}, {Aller}, {Bach},
  {Ibrahimov}, {Kovalev}, {Kurtanidze}, {Larionov}, {Lee}, {Leto},
  {L{\"a}hteenm{\"a}ki}, {Nilsson}, {Pursimo}, {Ros}, {Sumitomo}, {Volvach},
  {Aller}, {Arai}, {Buemi}, {Coloma}, {Doroshenko}, {Efimov}, {Fuhrmann},
  {Hagen-Thorn}, {Kamada}, {Katsuura}, {Konstantinova}, {Kopatskaya}, {Kotaka},
  {Kovalev}, {Kurosaki}, {Lanteri}, {Larionova}, {Mingaliev}, {Mizoguchi},
  {Nakamura}, {Nikolashvili}, {Nishiyama}, {Sadakane}, {Sergeev}, {Sigua},
  {Sillanp{\"a}{\"a}}, {Smart}, {Takalo}, {Tanaka}, {Tornikoski}, {Trigilio},
  \& {Umana}}]{vil07}
{Villata}, M., {Raiteri}, C.~M., {Aller}, M.~F., {et~al.} 2007, \aap, 464, L5

\bibitem[{{Villata} {et~al.}(2006){Villata}, {Raiteri}, {Balonek}, {Aller},
  {Jorstad}, {Kurtanidze}, {Nicastro}, {Nilsson}, {Aller}, {Arai}, {Arkharov},
  {Bach}, {Ben{\'{\i}}tez}, {Berdyugin}, {Buemi}, {B{\"o}ttcher}, {Carosati},
  {Casas}, {Caulet}, {Chen}, {Chiang}, {Chou}, {Ciprini}, {Coloma}, {di Rico},
  {D{\'{\i}}az}, {Efimova}, {Forsyth}, {Frasca}, {Fuhrmann}, {Gadway}, {Gupta},
  {Hagen-Thorn}, {Harvey}, {Heidt}, {Hernandez-Toledo}, {Hroch}, {Hu}, {Hudec},
  {Ibrahimov}, {Imada}, {Kamata}, {Kato}, {Katsuura}, {Konstantinova},
  {Kopatskaya}, {Kotaka}, {Kovalev}, {Kovalev}, {Krichbaum}, {Kubota},
  {Kurosaki}, {Lanteri}, {Larionov}, {Larionova}, {Laurikainen}, {Lee}, {Leto},
  {L{\"a}hteenm{\"a}ki}, {L{\'o}pez-Cruz}, {Marilli}, {Marscher}, {McHardy},
  {Mondal}, {Mullan}, {Napoleone}, {Nikolashvili}, {Ohlert}, {Postnikov},
  {Pursimo}, {Ragni}, {Ros}, {Sadakane}, {Sadun}, {Savolainen}, {Sergeeva},
  {Sigua}, {Sillanp{\"a}{\"a}}, {Sixtova}, {Sumitomo}, {Takalo},
  {Ter{\"a}sranta}, {Tornikoski}, {Trigilio}, {Umana}, {Volvach}, {Voss}, \&
  {Wortel}}]{vil06}
{Villata}, M., {Raiteri}, C.~M., {Balonek}, T.~J., {et~al.} 2006, \aap, 453,
  817

\bibitem[{{Villata} {et~al.}(2004{\natexlab{b}}){Villata}, {Raiteri},
  {Kurtanidze}, {Nikolashvili}, {Ibrahimov}, {Papadakis}, {Tosti}, {Hroch},
  {Takalo}, {Sillanp{\"a}{\"a}}, {Hagen-Thorn}, {Larionov}, {Schwartz},
  {Basler}, {Brown}, {Balonek}, {Ben{\'{\i}}tez}, {Ram{\'{\i}}rez}, {Sadun},
  {Boltwood}, {Carini}, {Barnaby}, {Coloma}, {Ros}, {Dai}, {Xie}, {Mattox},
  {Rodriguez}, {Asfandiyarov}, {Atkerson}, {Beem}, {Bloom}, {Chanturiya},
  {Ciprini}, {Crapanzano}, {de Diego}, {Efimova}, {Gardiol}, {Guerra},
  {Kahharov}, {Kapanadze}, {Karttunen}, {Kato}, {Kimeridze}, {Kudryavtseva},
  {Lainela}, {Lanteri}, {Larionova}, {Maesano}, {Marchili}, {Massone},
  {Monroe}, {Montagni}, {Nesci}, {Nilsson}, {Noble}, {Nucciarelli}, {Ostorero},
  {Papamastorakis}, {Pasanen}, {Peters}, {Pursimo}, {Reig}, {Ryle}, {Sclavi},
  {Sigua}, {Uemura}, \& {Wills}}]{vil04a}
{Villata}, M., {Raiteri}, C.~M., {Kurtanidze}, O.~M., {et~al.}
  2004{\natexlab{b}}, \aap, 421, 103

\bibitem[{{Villata} {et~al.}(1998){Villata}, {Raiteri}, {Lanteri}, {Sobrito},
  \& {Cavallone}}]{vil98b}
{Villata}, M., {Raiteri}, C.~M., {Lanteri}, L., {Sobrito}, G., \& {Cavallone},
  M. 1998, \aaps, 130, 305

\bibitem[{{Wagner} {et~al.}(1996){Wagner}, {Witzel}, {Heidt}, {Krichbaum},
  {Qian}, {Quirrenbach}, {Wegner}, {Aller}, {Aller}, {Anton}, {Appenzeller},
  {Eckart}, {Kraus}, {Naundorf}, {Kneer}, {Steffen}, \& {Zensusj}}]{wag96}
{Wagner}, S.~J., {Witzel}, A., {Heidt}, J., {et~al.} 1996, \aj, 111, 2187

\end{thebibliography}
\end{document}